\documentclass[technote]{IEEEtran}

\usepackage{cite,graphicx,subfigure,amsmath,amsfonts}

\usepackage{hyperref}

\newlength{\Lpr}
\newsavebox{\Bpr}

\newcommand{\D}[1]{\ensuremath{\displaystyle #1}}

\newcommand{\V}[1]{\mbox{\boldmath$\mathbf{#1}$\unboldmath}}

\newcommand{\sinc}{\ensuremath{{\mathrm{sinc}}}}

\newcommand{\bdm}{\begin{displaymath}}
\newcommand{\edm}{\end{displaymath}}

\newcommand{\be}[1]{\begin{equation} \label{#1}}
\newcommand{\ee}{\end{equation}}

\newcommand{\bae}[3]{
\begin{equation} \label{#1}
\renewcommand{\arraystretch}{#2}
\begin{array}{#3}}

\newcommand{\eae}{\end{array}\end{equation}}

\newcommand{\baen}[2]{
\begin{displaymath} 
\renewcommand{\arraystretch}{#1}
\begin{array}{#2}}

\newcommand{\eaen}{\end{array}\end{displaymath}}

\newcommand{\DefLetter}[4]{
\newcommand{#1}{\ensuremath{\V{#2}}} 
\newcommand{#3}{\ensuremath{\V{#4}}} 
}

\DefLetter{\vzer}{0}{\mzer}{0}
\DefLetter{\vone}{1}{\mone}{1}
\DefLetter{\va}{a}{\ma}{A}
\DefLetter{\vb}{b}{\mb}{B}
\DefLetter{\vc}{c}{\mc}{C}
\DefLetter{\vd}{d}{\md}{D}
\DefLetter{\ve}{e}{\me}{E}
\DefLetter{\vf}{f}{\mf}{F}
\DefLetter{\vg}{g}{\mg}{G}
\DefLetter{\vh}{h}{\mh}{H}
\DefLetter{\vi}{i}{\mi}{I}
\DefLetter{\vj}{j}{\mj}{J}
\DefLetter{\vk}{k}{\mk}{K}
\DefLetter{\vl}{l}{\ml}{L}
\DefLetter{\vm}{m}{\mm}{M}
\DefLetter{\vn}{n}{\mn}{N}
\DefLetter{\vpr}{p}{\mpr}{P}
\DefLetter{\vq}{q}{\mq}{Q}
\DefLetter{\vr}{r}{\mr}{R}
\DefLetter{\vs}{s}{\ms}{S}
\DefLetter{\vt}{t}{\mt}{T}
\DefLetter{\vur}{u}{\mur}{U}
\DefLetter{\vv}{v}{\mv}{V}
\DefLetter{\vw}{w}{\mw}{W}
\DefLetter{\vx}{x}{\mx}{X}
\DefLetter{\vy}{y}{\my}{Y}
\DefLetter{\vz}{z}{\mz}{Z}

\DefLetter{\vdel}{\delta}{\mdel}{\Delta}
\DefLetter{\vphi}{\phi}{\mphi}{\Phi}
\DefLetter{\vpsi}{\psi}{\mpsi}{\Psi}
\DefLetter{\vrho}{\rho}{\mrho}{\Lambda}
\DefLetter{\vxi}{\xi}{\mxi}{\Xi}

\DefLetter{\valpha}{\alpha}{\malpha}{\Alpha}
\DefLetter{\vbeta}{\beta}{\mbeta}{\Beta}
\DefLetter{\vlam}{\lambda}{\mlam}{\Lambda}
\DefLetter{\vsig}{\sigma}{\msig}{\Sigma}
\DefLetter{\vtau}{\tau}{\mtau}{\tau}
\DefLetter{\vtheta}{\theta}{\mtheta}{\Theta}
\DefLetter{\vome}{\omega}{\mome}{\Omega}
\DefLetter{\vzero}{0}{\mzero}{0}
\DefLetter{\vgam}{\gamma}{\mgam}{\Gamma}
\DefLetter{\veps}{\epsilon}{\meps}{\Epsilon}
\DefLetter{\veta}{\eta}{\meta}{\Eta}

\newcommand{\eqand}{\:\:\mbox{and}\:\:}

\newcommand{\DefFuncLetter}[2]{
\newcommand{#1}{\ensuremath{{\mathrm{#2}}}} 
}

\DefFuncLetter{\Fzer}{0}
\DefFuncLetter{\Fa}{a}
\DefFuncLetter{\FA}{A}
\DefFuncLetter{\Fb}{b}
\DefFuncLetter{\Fc}{c}
\DefFuncLetter{\FC}{C}
\DefFuncLetter{\Fd}{d}
\DefFuncLetter{\Fe}{e}
\DefFuncLetter{\Ff}{f}
\DefFuncLetter{\Fg}{g}
\DefFuncLetter{\FG}{G}
\DefFuncLetter{\Fh}{h}
\DefFuncLetter{\FH}{H}
\DefFuncLetter{\Fi}{i}
\DefFuncLetter{\Fk}{k}
\DefFuncLetter{\Fl}{l}
\DefFuncLetter{\FL}{L}
\DefFuncLetter{\Fm}{m}
\DefFuncLetter{\Fn}{n}
\DefFuncLetter{\Fnr}{n}
\DefFuncLetter{\FN}{N}
\DefFuncLetter{\Fo}{o}
\DefFuncLetter{\FO}{O}
\DefFuncLetter{\Fpr}{p}
\DefFuncLetter{\FPr}{P}
\DefFuncLetter{\Fq}{q}
\DefFuncLetter{\Fr}{r}
\DefFuncLetter{\Fs}{s}
\DefFuncLetter{\FS}{S}
\DefFuncLetter{\Ft}{t}
\DefFuncLetter{\FT}{T}
\DefFuncLetter{\Fu}{u}
\DefFuncLetter{\FU}{U}
\DefFuncLetter{\Fv}{v}
\DefFuncLetter{\Fw}{w}
\DefFuncLetter{\FW}{W}
\DefFuncLetter{\Fx}{x}
\DefFuncLetter{\Fy}{y}
\DefFuncLetter{\FY}{Y}
\DefFuncLetter{\Fz}{z}
\DefFuncLetter{\FZ}{Z}
\DefFuncLetter{\Falp}{\alpha}
\DefFuncLetter{\Fbet}{\beta}
\DefFuncLetter{\Fchi}{\chi}
\DefFuncLetter{\Fdel}{\delta}
\DefFuncLetter{\Fzet}{\zeta}
\DefFuncLetter{\Feta}{\eta}
\DefFuncLetter{\Fphi}{\phi}
\DefFuncLetter{\FPhi}{\Phi}
\DefFuncLetter{\Fpsi}{\psi}
\DefFuncLetter{\FPsi}{\Psi}
\DefFuncLetter{\Fgam}{\gamma}
\DefFuncLetter{\FGam}{\Gamma}
\DefFuncLetter{\Flam}{\lambda}
\DefFuncLetter{\Fsig}{\sigma}
\DefFuncLetter{\Ftau}{\tau}
\DefFuncLetter{\Fome}{\omega}
\DefFuncLetter{\Feps}{\epsilon}
\DefFuncLetter{\Fthe}{\theta}
\DefFuncLetter{\Fvar}{\vartheta}

\DefFuncLetter{\FB}{B}
\DefFuncLetter{\FD}{D}
\DefFuncLetter{\FE}{E}
\DefFuncLetter{\FF}{F}
\DefFuncLetter{\FI}{I}
\DefFuncLetter{\FJ}{J}
\DefFuncLetter{\FM}{M}
\DefFuncLetter{\FR}{R}
\DefFuncLetter{\FX}{X}

\newcommand{\DefCalLetter}[2]{
\newcommand{#1}{\ensuremath{\mathcal{#2}}} 
}
\DefCalLetter{\CC}{C}
\DefCalLetter{\CD}{D}

\DefCalLetter{\CS}{S}
\DefCalLetter{\CV}{V}

\newcommand{\DefSubLetter}[2]{
\newcommand{#1}{\mathrm{#2}} 
}

\DefSubLetter{\slzer}{0}
\DefSubLetter{\sla}{a}
\DefSubLetter{\slA}{A}
\DefSubLetter{\slb}{b}
\DefSubLetter{\slB}{B}
\DefSubLetter{\slc}{c}
\DefSubLetter{\slC}{C}
\DefSubLetter{\sld}{d}
\DefSubLetter{\slD}{D}
\DefSubLetter{\sle}{e}
\DefSubLetter{\slE}{E}
\DefSubLetter{\slf}{f}
\DefSubLetter{\slF}{F}
\DefSubLetter{\slg}{g}
\DefSubLetter{\slG}{G}
\DefSubLetter{\slh}{h}
\DefSubLetter{\slH}{H}
\DefSubLetter{\sli}{i}
\DefSubLetter{\slI}{I}
\DefSubLetter{\slk}{k}
\DefSubLetter{\sll}{l}
\DefSubLetter{\slL}{L}
\DefSubLetter{\slm}{m}
\DefSubLetter{\slM}{M}
\DefSubLetter{\sln}{n}
\DefSubLetter{\slnr}{n}
\DefSubLetter{\slN}{N}
\DefSubLetter{\slo}{o}
\DefSubLetter{\slp}{p}
\DefSubLetter{\slP}{P}
\DefSubLetter{\slq}{q}
\DefSubLetter{\slQ}{Q}
\DefSubLetter{\slr}{r}
\DefSubLetter{\slR}{R}
\DefSubLetter{\sls}{s}
\DefSubLetter{\slS}{S}
\DefSubLetter{\slt}{t}
\DefSubLetter{\slT}{T}
\DefSubLetter{\slu}{u}
\DefSubLetter{\slU}{U}
\DefSubLetter{\slv}{v}
\DefSubLetter{\slw}{w}
\DefSubLetter{\slW}{W}
\DefSubLetter{\slx}{x}
\DefSubLetter{\slX}{X}
\DefSubLetter{\sly}{y}
\DefSubLetter{\slY}{Y}
\DefSubLetter{\slz}{z}
\DefSubLetter{\slZ}{Z}

\DefSubLetter{\slalp}{\alpha}
\DefSubLetter{\slbet}{\beta}
\DefSubLetter{\sldel}{\delta}
\DefSubLetter{\slDel}{\Delta}
\DefSubLetter{\sleps}{\epsilon}
\DefSubLetter{\slgam}{\gamma}
\DefSubLetter{\slphi}{\phi}
\DefSubLetter{\sltau}{\tau}
\DefSubLetter{\slxi}{\xi}
\DefSubLetter{\slthe}{\theta}

\begin{document}

\title{Efficient Sampling of Band-limited Signals from Sine Wave Crossings}

\author{J. Selva  \thanks{Copyright  (c) 2010  IEEE.  Personal  use  of this   material is
    permitted. However,  permission to use this  material for any  other purposes  must be
    obtained from the IEEE by sending a request  to pubs-permissions@ieee.org.  The author
    is with  the  Dept.  of  Physics, Systems   Engineering  and Signal  Theory  (DFISTS),
    University    of    Alicante,  P.O.Box    99,   E-03080    Alicante,  Spain   (e-mail:
    jesus.selva@ua.es).  This work has been supported by the Spanish Ministry of Education
    and Science (MEC), Generalitat Valenciana (GV), and by the University of Alicante (UA)
    under    the  following  projects/programmes:   TEC2005-06863-C02-02,  HA2007-075  and
    ``Ram\'{o}n y Cajal'' (MEC); ACOMP07-087 and GV07/214 (GV); and GRE074P (UA).  }}

\maketitle

\markboth{}{}

\begin{abstract}

  This correspondence presents  an  efficient method   for reconstructing  a  band-limited
  signal in the discrete domain from its crossings  with a sine wave.  The method makes it
  possible to design A/D converters that only deliver the crossing timings, which are then
  used to interpolate the input  signal at arbitrary  instants.  Potentially, it may allow
  for    reductions in  power  consumption   and complexity   in   these converters.   The
  reconstruction  in the discrete  domain is based  on a recently-proposed modification of
  the Lagrange interpolator,  which is readily  implementable  with linear complexity  and
  efficiently, given that it  re-uses known schemes   for variable fractional-delay  (VFD)
  filters.  As  a spin-off, the method  allows one to  perform spectral analysis from sine
  wave  crossings   with the  complexity   of the  FFT.    Finally,  the  results   in the
  correspondence are validated in several numerical examples.

\end{abstract}

\section{Introduction}

The analog-to-discrete (A/D) conversion is the first step for the discrete-time processing
of continuous signals. This  conversion is fundamentally  based  on the Sampling  Theorem,
which states that a band-limited signal can be recovered from its regularly-spaced samples
taken at least at twice the Nyquist  rate.  However, some  authors early noticed that this
recovery is also  possible from the  signal's zeros, or  from  its crossings with  another
signal like a sine wave, \cite{Voelcker73,Voelcker73b,BarDavid74,Requicha80}.  This is due
to the fact that  a band-limited signal  is an entire  function  of exponential type,  for
which there is a   factorization in terms   of its roots akin    to that of   conventional
polynomials, (Hadamard's factorization  theorem \cite[chapter 2]{Boas54}).  A  consequence
of this is that it would be possible, in principle, to design  A/D converters in which the
sample  quantization is substituted by  a zero crossing  detector and  an accurate timing,
\cite{Kay86}.  This new procedure would eliminate the need to  quantize any signal samples,
so decreasing the complexity and power consumption of A/D converters, provided there is an
accurate timing available.  Besides, it would  mainly re-use existing technologies,  given
that zero crossing detection is implicit in many existing systems.  This last point can be
readily seen in the current trend in A/D converter design,  in which the sample amplitudes
are  turned into  zero crossings,  which  can then be  accurately  detected with low-power
consumption, \cite{Lee10,Lee07}.

The main obstacle for this alternative procedure is how the signal should be reconstructed
or processed  in the  discrete  domain, since  Hadamard's  factorization theorem  does not
directly lead to efficient implementations,  due to its  slow convergence rate.  Here, the
usual approach in the literature consists in approximating the signal in a finite interval
using a trigonometric  polynomial,   \cite{Kay86,Sreenivas92,Kumaresan10}.  But then   the
interpolation error decreases  only as $\FO(1/N)$,  while the complexity per  interpolated
value is $\FO(N)$, where $N$ is the number of crossings inside the  interval.  So, in this
approach it is necessary to employ a large $N$ to ensure  an acceptable accuracy, with the
associated high complexity.

The purpose of this correspondence is to present a method for  overcoming this obstacle, that makes
it possible   to  reconstruct the    bandlimited  signal from   its  sine  wave  crossings
efficiently.    The method  is   based on  viewing  the  reconstruction  as  a problem  of
interpolation from nonuniform samples, to which the  efficient technique in \cite{Selva09}
is applied. Relative to the state of the art, it has several advantages:

\begin{itemize}
\item  The       complexity is  reduced      significantly.   If  in     the    approach in
  \cite{Kay86,Sreenivas92,Kumaresan10}, a complexity   $\FO(N)$ (per interpolated   value)
  gives an interpolation error $\FO(1/N)$, with the proposed  method a complexity $\FO(N)$
  gives an error $\FO(\Fe^{-\pi (1-BT) N})$, where $B$ is the signal's two-sided bandwidth
  and $T$ is the average crossing separation. In practice this means that ``any'' accuracy
  can be achieved with a small $N$.

\item The method is based on the evaluation of a fixed smooth function and on the Lagrange
  interpolator.  Besides, it can be  evaluated with cost  $\FO(N)$ per interpolated value,
  and can be implemented by re-using efficient designs for variable fractional-delay (VFD)
  filters, \cite[Sec. IV]{Selva09}.

\item As a spin-off,  the method permits  one to perform  spectral analysis from sine wave
  crossings   with complexity  $\FO(N\log  N)$,  while  the  usual  method has  complexity
  $\FO(N^2)$, \cite[Sec. IV]{Kay86}.
\end{itemize} 

The correspondence has been organized as follows.   The next section  reviews the state of
the art and presents the problem formulation.  In it, it is  shown that the reconstruction
from sine wave  crossings can  be  turned into  an  interpolation problem from  nonuniform
samples.  Then, this last problem is  addressed in Sec.   \ref{sec:ps}, where the solution
adopted  is  that  in   the  recent reference    \cite{Selva09},  based on  the   Lagrange
interpolator.   The main result is  a simple interpolation  formula for reconstructing the
signal from its sine wave crossings.   Then, Sec.  \ref{sec:ses}  addresses the problem of
analyzing the spectrum from   sine wave crossings   in the light of   the formula in  Sec.
\ref{sec:ps}.  It turns out that this formula makes it possible  to reduce the complexity
from  the usual $\FO(N^2)$ order   to order $\FO(N\log  N)$.   Finally, Sec.  \ref{sec:ne}
validates the results in the correspondence through a numerical example.

\section{State of the art and problem formulation}
\label{sec:sap}

The usual  representation for a  real finite-energy signal  $\Fs(t)$  with spectrum inside
$[-B/2,B/2]$ is given by the Sampling  Theorem. If the samples  $\Fs(n/B)$ are known, this
theorem states that $\Fs(t)$ can be perfectly reconstructed using the series
\be{eq:1}
\Fs(n/B+u)=\sum_{p=-\infty}^{\infty}\Fs((n-p)/B)\sinc(p+Bu),
\ee
where $n$ is an integer and $u$ is any time shift following  $-1/(2B)\leq u<1/(2B)$.  This
series is the basis of  most processing algorithms for  continuous signals in the discrete
domain.   Several   authors  \cite{Voelcker73,Voelcker73b,Requicha80} soon    noticed that
$\Fs(t)$ can alternatively be viewed as a \emph{polynomial of  infinite degree}, which can
be described in terms of its roots.  The reason  why is that  a band-limited signal can be
regarded  as an analytic function  over the whole  complex plane, if  the  $t$ variable is
allowed  to take complex  values.  Besides, this kind  of function is  bounded on the real
axis (real  $t$), and its  maximum growth  rate is  that  of $\Fe^{\pi  B |t|}$  along the
imaginary axis.  It can be shown  that this kind  of signal admits the
representation
\be{eq:4}
\Fs(t)=K\lim_{R\rightarrow \infty}\sideset{}{'}\prod_{|\tau_p|\leq R} (1-t/\tau_p),
\ee
where  $K$ is  a real
constant, and the $\tau_p$ are  complex roots  which  appear in  conjugate pairs,
$\tau_p\neq 0$,  and may not  be distinct, \cite[Theorem  VI]{Titchmarsh26}. The prime (')
means that the factor should be replaced with $t$ if $t_p=0$ %
\footnote{The  factorization    in (\ref{eq:4})    is   due  to   Titmarsch  \cite[Theorem
  VI]{Titchmarsh26}, and is a refinement of Hadamard's factorization theorem \cite[chapter
  2]{Boas54} for $\Fs(t)$ bounded on the real axis. Note that there is a bug in Eq. [4] of
  \cite{Selva09}. This last equation is Hadamard's factorization and its correct form is
\begin{displaymath}
\Fphi(t)\equiv A\Fe^{-t\sum_n 1/t_n}
\sideset{}{'}\prod_{n=-\infty}^{\infty} (1-t/t_n)\Fe^{t/t_n}.
\end{displaymath}
 }. 
Eq.  (\ref{eq:4}) is   the explicit root    factorization of $\Fs(t)$ as   infinite-degree
polynomial, and from it it is obvious that the roots $\tau_p$  determine the signal except
for the scale factor $K$. The root density in any circle  $|t|<\tau$ follows the same rule
as the sample density in  (\ref{eq:1}), i.e, the circle  $|t|<\tau$ contains either $2\tau
B$ samples in (\ref{eq:1}) or $2\tau B$ roots in (\ref{eq:4}) asymptotically.

The factorization in  Eq.  (\ref{eq:4}) suggested the  possibility of performing the usual
processing of continuous  signals in the discrete  domain using the   zeros of the  signal
instead  of its samples.   In such approach, the  analog-to-discrete (A/D) conversion would
consist in acquiring the signal's zeros, and the discrete-to-analog (D/A) conversion would
be based on the evaluation of a formula like (\ref{eq:4}).  However,  it was soon realized
that there were two main problems,  \cite{Requicha80,Kay86}.  The first  was how the roots
should  be located efficiently   in the A/D   conversion, since they   may have a non-null
imaginary  part.  And the  second was how  the infinite product  in (\ref{eq:4}) should be
approximated  so  as to implement   the D/A conversion,  since  in practice  only a finite
sequence of roots $\tau_p$ is known, and (\ref{eq:4}) converges slowly.

A solution for the first problem was readily found \cite[Sec V]{Requicha80}, and consisted
in subtracting a sinusoidal $A_s \sin(\pi B t)$ to $\Fs(t)$, where
\be{eq:5}
A_s=\sup_{t\;\mathrm{real}} |\Fs(t)|. 
\ee
This  simple procedure  actually solved  the problem  of locating  the  roots, because the
subtraction of this sine wave ``moves''  all roots to the  real axis, due  to a theorem of
Duffin and Schaeffer,  \cite{Duffin38}.  Specifically,  $\Fs(t)-A_s\sin(\pi  B t)$ can  only
have zeros on the real axis, and each of them can be viewed as a zero of $A\sin(\pi B t)$,
which has   been   shifted by  at    most $1/(2B)$.  So,     in notation, all    zeros  of
$\Fs(t)-A_s\sin(\pi B  t)$  have the  form   $n/B+\delta_n$, where $n$   is  an integer  and
$|\delta_n|\leq   1/(2B)$, and two     consecutive zeros may     overlap only at  instants
$n/B+1/(2B)$.   Using   this description,   the   factorization in   Eq.  (\ref{eq:4}) for
$\Fs(t)-A_s\sin(\pi B t)$ is
\be{eq:6}
\Fs(t)-A_s\sin(\pi B t)=K' (t_0-t)
\prod_{k=1}^\infty \Big(1-\frac{t}{t_{-k}}\Big)
\Big(1-\frac{t}{t_k}\Big),
\ee
where $K'$ is a real  constant and
\be{eq:21}
t_n\equiv n/B+\delta_n.
\ee
From Eq. (\ref{eq:6}), the basic design for the desired A/D converter was clear.  First an
oscillator would be used   to generate the   wave $A_s\sin(\pi t/B)$,   which would then  be
subtracted from $\Fs(t)$.  Afterward, the zero crossings would  be detected using a gate,
and the converter output would be the sequence of shifts $\delta_n$ in Eq. (\ref{eq:21}).

As to the second problem, the usual solution to date  consists in approximating the signal
using a trigonometric polynomial, \cite{Kay86}. In short, if  the finite sequence of roots
$t_n, t_{n+1}, \ldots, t_{n+M-1}$ is known, then $\Fs(t)-A_s\sin(\pi B t)$ is interpolated
using a trigonometric polynomial  of order $M$ which  is zero at  $t_{n+m}$, $0\leq m <M$.
However, this solution  is not satisfactory  since its accuracy is  poor even for a  large
number of  roots.  This  complexity issue  is  the main  obstacle for  achieving efficient
implementations.

The purpose of this  correspondence is to  provide an efficient   solution to this  second
problem.  The key point is to realize that approximating $\Fs(t)$ from  the roots $t_n$ of
$\Fs(t)-A_s\sin(\pi B t)$  is the same  as approximating $\Fs(t)$ from  its value at these
instants, since  $\Fs(t_n)=A_s\sin(\pi  B t_n)$.    So, this  is   actually a  problem  of
interpolation from nonuniform samples.

\section{Proposed solution}
\label{sec:ps}

Reference \cite{Selva09} presents an efficient  and accurate interpolator for band-limited
signals from nonuniform  samples,  applicable to  signals  of the  same type  as $\Fs(t)$,
(bandwidth $B$, supremum amplitude $A_s$). Consider one such signal $\Fz(t)$ and assume it
is known at a set of instants $\tau_p$ of the form
\be{eq:61}
\tau_p\equiv pT+\eta_p, \; -P\leq p\leq P,
\ee
where  the  period $T$  and  the shifts  $\eta_p$ follow  $BT<1$ and  $\eta_p<T/2$ (strict
inequalities).  The   approach  in  \cite{Selva09} consists    in  applying the   Lagrange
interpolator  to the product  $\Fz(t)\Fgam(t)$, where $\Fgam(t)$ is  a fixed function, and
then solving for the value of $\Fz(t)$.  The interpolator's formula is
\be{eq:51}
\Fz(t)\approx\frac{1}{\Fgam(t)}
\sum_{p=-P}^{P} \Fz(\tau_p)\Fgam(\tau_p)\frac{\FL(t)}{\FL'(\tau_p)(t-\tau_p)},
\ee
where 
\be{eq:52}
\FL(t)\equiv \prod_{p=-P}^P t-\tau_p,
\ee
and $\Fgam(t)$ is   given in Ap.   \ref{ap:wfl}  and only depends   on $B$, $T$  and  $P$.
Usually, Eq. (\ref{eq:51}) is applied for $|t|\leq T/2$, though  wider ranges are allowed.
In \cite{Selva09}, it was shown that the accuracy of (\ref{eq:51}) increases exponentially
with $P$. So, in practice a small $P$ is enough to obtain high accuracy.

To assess the  accuracy of (\ref{eq:51}),  assume $BT=0.7$  and  $|\Fz(t)|\leq 1$.   Then,
following the  analysis   in  \cite[Sec.  III]{Selva09},    the interpolation  error    of
(\ref{eq:51}) for the function $\Fgam(t)$ specified in Ap. \ref{ap:wfl} is well fitted by
\bae{eq:34}{1.5}{l}
\D{
\epsilon\, \mathrm{(dB)}\approx 4.12106+66.6044\,\delta-9.35838\,\delta^2}\hspace{1cm}{}\\
{}\hfill\D{  -8.30873P+3.13419\, \delta P-0.125803\, \delta^2 P },
\eae
where $\delta$ is a bound on the deviations  of the instants $\tau_p$  from a uniform grid
with  spacing $T$  and  $\delta<T/2$.   So, if   $\delta=T/4$ the  value  $P=10$  gives an
interpolation error below $\epsilon= -55$ dB, and $P=16$ gives an  error below $\epsilon =
-100$ dB.  Any  practical accuracy can be  obtained by slightly  increasing $P$ for  fixed
$\delta$. (For a detailed analysis, see the previous reference.)

Coming back to the problem in the previous section, the sine wave crossings are equivalent
to  nonuniform samples  of a  form  similar to  that  in (\ref{eq:51}). Specifically,  the
crossing instants  in  (\ref{eq:21}) have  a  maximum deviation   from the  uniform  grid,
$|\delta_n|\leq 1/(2B)$, and the sample values are given by the sine wave, since
\be{eq:54}
\Fs(t_n) = A_s\sin(\pi B t_n) = A_s(-1)^n\sin(\pi B \delta_n).
\ee
However  Eq.  (\ref{eq:51}) is  not  applicable  to  $\Fs(t)$ since   the regular  grid in
(\ref{eq:21}) matches exactly the Nyquist rate and it may be $|\delta_n|=1/(2B)$, while in
(\ref{eq:61}) there     is  some sampling   inefficiency,   $BT<1$,  and   it    is always
$|\eta_p|<T/2$.  These latter conditions  can be easily imposed  on the sampling scheme by
slightly increasing the amplitude  and frequency of  the sine wave.   So if $A_s\sin(\pi B
t)$ in  (\ref{eq:6}) is replaced  by another  sine  wave $A\sin(\pi t/T)$,  with amplitude
$A>A_s$ and semi-period fulfilling $T>0$ and  $BT<1$, the theorem  of Duffin and Schaeffer
in the previous section ensures that the only zeros of the signal
\be{eq:58}
\Fs(t)-A\sin(\pi t/T)
\ee
are simple and occur at positions
\be{eq:60}
t_n=nT+\delta_n
\ee
for   integer $n$.  Besides, the   shifts  $\delta_n$ now   follow  the  strict inequality
$|\delta_n|\leq\delta<T/2$, where
\be{eq:59}
\delta\equiv (T/\pi)\arcsin(A_s/A),
\ee
since there can be no zero crossings whenever $A|\sin(\pi t/T)| >A_s$.

\begin{figure}
\begin{center}
\includegraphics{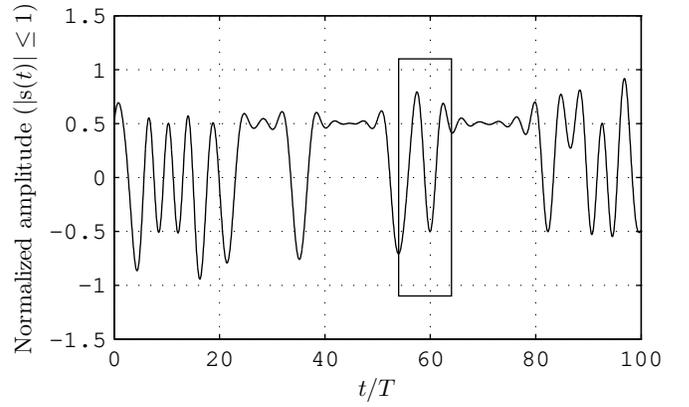}
\end{center}
\caption{\label{fig:4} Random BPSK signal with bandwidth $B=0.7/T$ and peak amplitude
  1. The symbols used to generate this signal were random $\pm 1$ values, and the
  modulating pulse was a raised cosine with roll-off factor 0.2. The signal's peak
  amplitude was scaled to one.}
\end{figure}

\begin{figure}
\begin{center}
\includegraphics{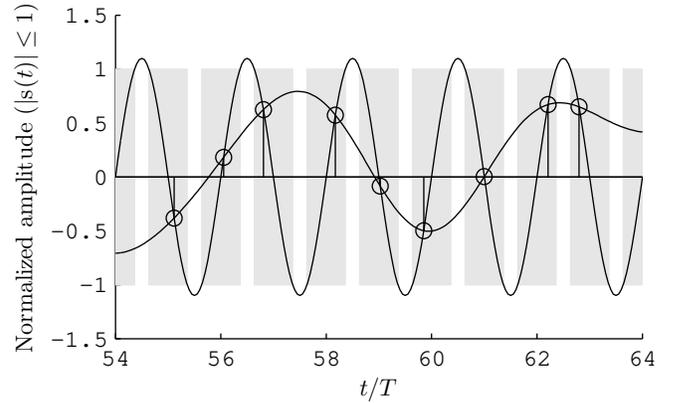}
\end{center}
\caption{\label{fig:1} Short piece of the BPSK signal in Fig. \ref{fig:4}, overlapped with
  the sine wave  $A\sin(\pi t/T)$, with  $A=1.1$. The stems indicate  the positions of the
  crossings with the sine wave. These can only take place inside the rectangles, and there
  is exactly one in each of them. }
\end{figure}

To check these conditions numerically, consider the BPSK signal in Fig.  \ref{fig:4}.
This  signal was generated   by modulating a   raised cosine pulse   of  roll-off 0.2  and
bandwidth $B=0.7/T$  with a sequence of random amplitudes $\pm 1$.   Then the signal's peak
amplitude was scaled to  1.  Fig.  \ref{fig:1} shows  the zone marked  with a rectangle in
Fig.  \ref{fig:4}, together with the sine wave $A\sin(\pi t/T)$.
The conditions in (\ref{eq:60}) and  (\ref{eq:59}) mean that  the crossings with the sine
wave can only take place inside the shaded rectangles, and there is exactly one in each of
them, as can be seen in this example.

Let us derive the final interpolation formula. Assume $\Fs(t)$ must  be interpolated at an
arbitrary $t$ from the crossings with $A\sin(\pi t/T)$, ($A>A_s$, $BT<1$).  Any $t$ can be
written uniquely in the form $t=nT+u$ with integer $n$ and $-T/2\leq u <T/2$.  Besides the
signal $\Fz(u)=\Fs(nT+u)$ with time variable $u$ is of the same  type as $\Fs(t)$, since a
time shift affects neither the bandwidth nor the supremum  amplitude.  So (\ref{eq:51}) is
valid on $\Fz(u)$ with instants $\tau_p=pT+\delta_{n+p}$ and values $\Fz(pT+\delta_{n+p})=
A(-1)^{n+p}\sin(\pi\delta_{n+p}/T)$, i.e, it is
\bae{eq:56}{1.5}{l}
\D{\Fs(nT+u)\approx\frac{1}{\Fgam(u)}}\cdot\\
\D{\sum_{p=-P}^{P} 
\frac{A(-1)^{n+p}\sin(\pi\delta_{n+p}/T)\Fgam(pT+\delta_{n+p})\FL_n(u)}
{\FL_n'(pT+\delta_{n+p})(u-pT-\delta_{n+p})}}\\
\D{=\frac{A(-1)^n}{\Fgam(u)}\cdot}\\
\D{\sum_{p=-P}^{P}\frac{(-1)^p\sin(\pi\delta_{n+p}/T)\Fgam(pT+\delta_{n+p})\FL_n(u)}
{\FL_n'(pT+\delta_{n+p})(u-pT-\delta_{n+p})}}
\eae
where
\be{eq:57}
\FL_n(u)\equiv \prod_{p=-P}^P u-pT-\delta_{n+p}.
\ee
Eq. (\ref{eq:56}) is the final interpolation formula. 

\begin{figure}
\subfigure[$P=2$]{\label{fig:2}\includegraphics{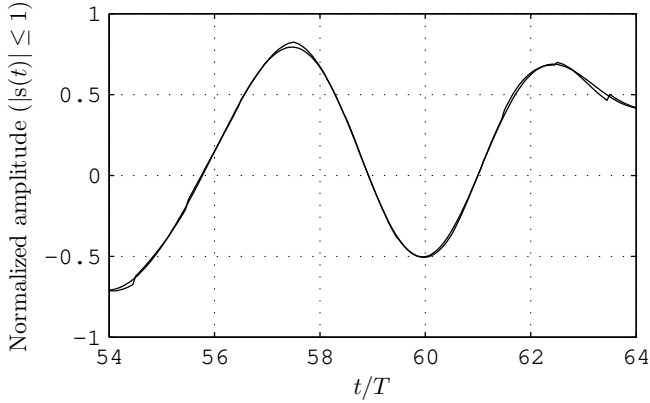}}
\subfigure[$P=3$]{\label{fig:3}\includegraphics{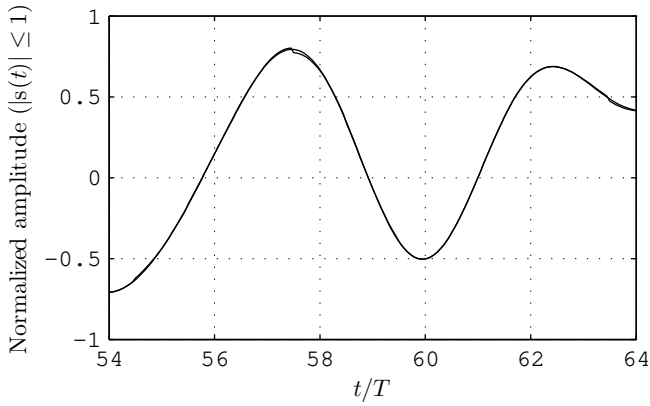}}
\caption{Sample BPSK signal in \ref{fig:1} and interpolated signal using (\ref{eq:56})
for $P=2,\,3$.}
\end{figure}
For the   example  in Fig.    \ref{fig:1}, Fig.   \ref{fig:2}  shows  the signal  and  its
interpolated   version  using  (\ref{eq:56})   for   $P=2$. The    discontinuities in  the
interpolated  signal take place at  $t=nT+T/2$, integer $n$, because  the set of crossings
used is different for  each $n$.  Notice however  that just five  crossings ($P=2$) give a
good  accuracy.  Fig.   \ref{fig:3}  shows  the same  comparison   but for  $P=3$,  (seven
crossings).   The    difference between  both   signals  is  much   smaller  than  in Fig.
\ref{fig:2}.  For $P>3$  the error becomes too  small to be represented  this way. For  an
error analysis see Sec. \ref{sec:ne}.
  
The formula in  Eq.  (\ref{eq:56}) yields  discrete-time processing methods for delivering
samples of  $\Fs(t)$ with any spacing,  simply by assigning proper  values to $n$ and $u$.
The simplest  case is for  spacing $T$, simply  by setting  $u=0$. For  a generic  grid of
instants $n_1T_1$  with  $T_1>0$ and  integer  $n_1$, the   grid  sample $\Fs(n_1T_1)$  is
obtained  from (\ref{eq:56}),  simply by  setting  $n$ and  $u$  equal  to  the modulo-$T$
decomposition of $n_1 T_1$ in (\ref{eq:56}), i.e,
\be{eq:29}
n=\lfloor n_1 T_1/T+1/2\rfloor\eqand u=n_1T_1-nT.
\ee

As to the  efficient implementation of  Eq.  (\ref{eq:56}), it   was shown in  \cite[Sec.
IV]{Selva09}  that it  can be evaluated  with cost just   $\FO(N)$. See also the
numerical examples in \cite[Sec. V]{Selva09}.

\section{Spectral estimation from sine wave crossings}
\label{sec:ses}

The formula  in (\ref{eq:56}) makes it  possible to interpolate   the input signal  at any
instant from its  crossings with the  sine wave, and  the cost of  this operation is  just
$\FO(-\log\epsilon)$, where $\epsilon$ is  a bound on   the interpolation error.  This  is
because the error of (\ref{eq:56}) decreases exponentially with trend $\Fe^{-\pi(1-BT)P}$.
So, if $\epsilon$ is set below the working numerical error,  Eq.  (\ref{eq:56}) allows one
to obtain one sample of $\Fs(t)$ with a small and constant computational cost.  Therefore,
the cost  of computing $N$  samples  in a  regular grid  with  arbitrary spacing  $T_1$ is
$\FO(N)$.  Once these samples are available, the  situation is the usual  one in which the
signal's spectrum is  estimated from regularly  spaced samples, and  any of the well-known
techniques in spectral analysis becomes  applicable, \cite{Kay88}.  Since these techniques
are based  on the FFT  whose complexity  is  $\FO(N\log N)$, it  is  clear that  the total
complexity is also $\FO(N\log N)$. A numerical example is presented in the next section.

\section{Numerical examples}
\label{sec:ne}

\subsection{Sampling of a BPSK signal}

To validate
the results in a specific example, a BPSK signal $\Fs(t)$ was generated with the following
parameters,
\begin{itemize}
\item Modulating pulse: raised cosine with roll-off $0.2$.
\item Random amplitudes equal to $\pm 1$.
\item Total two-sided bandwidth $B=0.7/T$.
\item Time interval $I=[0,(N-1)T]$ with $N=1024$.
\end{itemize}
\begin{figure}
\begin{center}
\includegraphics{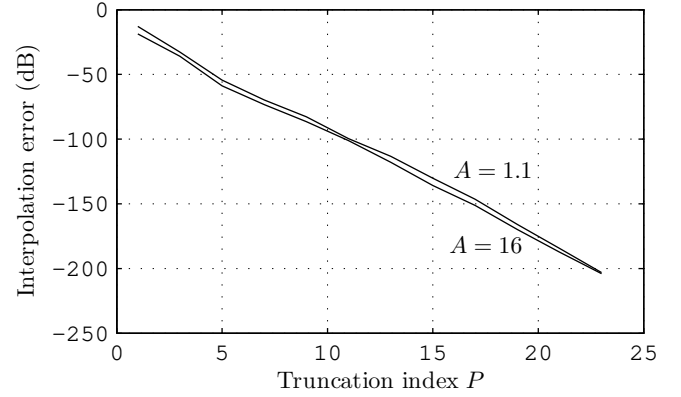}
\end{center}
\caption{\label{fig:6} Error in the interpolation of $\Fs(t)$ from its sine wave crossings
  versus index $P$. The error norm is the supremum $\sup_{t\in I} |\Fs(t)-\hat{\Fs}(t)|$
  where $\hat{\Fs}(t)$ is the interpolated signal.   }
\end{figure}
Then several numerical  experiments  were conducted.   

The  first  experiment consisted  in  interpolating $\Fs(t)$ in  $I$  from  its  sine wave
crossings.  The result is  shown in Fig.   \ref{fig:6} for $A=1.1$  and  $A=16$, where the
error norm is the maximum over $I$, that is, if $\hat{\Fs}(t)$ is the interpolated signal,
then the ordinate in this figure is
\be{eq:36}
\sup_{t\in I} |\Fs(t)-\hat{\Fs}(t)|.
\ee
Notice that this error decreases exponentially  with $P$. Besides,  the values of $\delta$
for $A=1.1$ and $A=16$ are $0.36T$ and $0.02T$, respectively, but the error is roughly the
same in both cases. So, the fact that the sampling instants may differ  from the grid $nT$
(integer $n$) has a minimal effect on the performance.

In the second experiment, a white noise process $\Fw(t)$ of bandwidth $B$ was added to
$\Fs(t)$. Then, $\Fs(t)+\Fw(t)$ was sampled at instants $nT$ (integer $n$) in $I$, and
these samples were also interpolated from the sine wave crossings of $\Fs(t)+\Fw(t)$ for
$A=3$ and $P=4$ and $9$. 
\begin{figure}
\begin{center}
\includegraphics{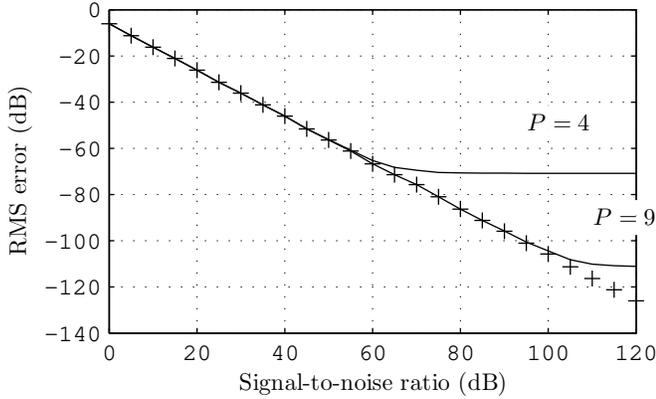}
\end{center}
\caption{\label{fig:5} RMS  error  versus the  SNR  in the  interpolation  of the  samples
  $\Fs(nT)$ from the sine wave crossings of the noisy realization $\Fs(t)+\Fw(t)$. }
\end{figure}
Fig. \ref{fig:5} shows the resulting root-mean-square (RMS)
error. The crosses ($+$) indicate the deviation of samples $\Fs(nT)+\Fw(nT)$.   
The other two curves are the RMS errors for $P=4$ and $P=9$, given by
\be{eq:37}
\Big(\frac{1}{N}\sum_{nT\in I} |\Fs(nT)-\hat{\Fs}_1(nT)|^2\Big)^{1/2},
\ee
where  $\Fs_1(t)$ is the  value interpolated from   the sine wave   crossings of the noisy
signal $\Fs(t)+\Fw(t)$. The curve for either value of $P$ overlaps the sample deviation up
to an SNR threshold which is fixed by the specific value of $P$.  So, below this threshold,
the performance  is the  same if  either the  signal  is directly  sampled,  or if  it  is
interpolated from its sine wave crossings. The threshold can  be fixed to  an SNR as large
as desired, simply  by slightly increasing  $P$, due to  the exponential dependence of the
interpolation error on $P$.

\begin{figure}
\begin{center}
\includegraphics{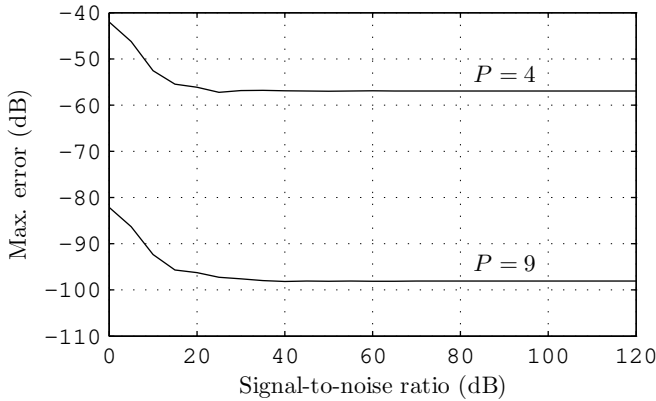}
\end{center}
\caption{\label{fig:7} Maximum difference between the samples $\Fs(nT)+\Fw(nT)$ and their
  interpolated values from sine wave crossings versus the SNR. The error norm if that
  defined in (\ref{eq:38}).}
\end{figure}

\subsection{Spectral analysis}

\begin{figure}
\begin{center}
\includegraphics{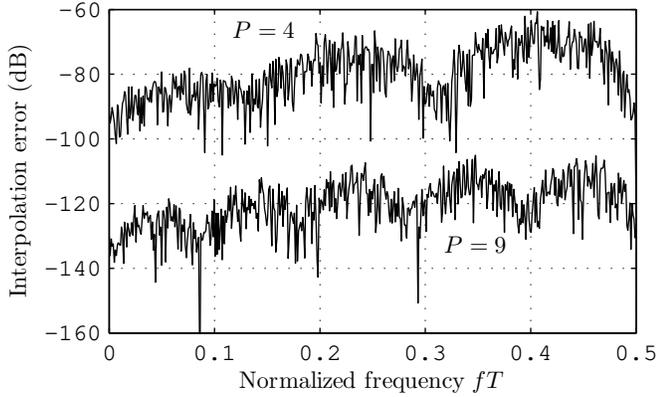}
\end{center}
\caption{\label{fig:9} Amplitude of the difference between the spectrum of
  $\Fs(nT)+\Fw(nT)$ and that of $\Fs_1(nT)$. }
\end{figure}

As to the spectral analysis, it is worth comparing the conventional procedure from uniform
samples, with the  one proposed in  this correspondence from sine  wave crossings. In  the
conventional   procedure, the samples   $\Fs(nT)+\Fw(nT)$  would be  delivered   by an A/D
converter, and then  any of the  existing spectral analysis  methods  would be applied  to
these data, \cite{Kay88}. And in the  proposed procedure, the  A/D converter would deliver
the sine wave crossing timings  $\delta_n$, then the uniform  samples $\Fs_1(nT)$ would be
computed using Eq.  (\ref{eq:56}), and finally the spectral analysis would  be the same as
in the  conventional procedure,  i.e,  it would  be  performed on the  samples $\Fs_1(nT)$
instead of $\Fs(nT)+\Fw(nT)$.   The fact is  that the result  of both procedures would  be
\emph{the same}  up to the  numerical accuracy in  use.  This can be  readily seen in Fig.
\ref{fig:7},   in which    the   error  measure is      the  maximum difference    between
$\Fs(nT)+\Fw(nT)$ and $\Fs_1(nT)$,
\be{eq:38}
\sup_{nT\in I} |\Fs(nT)+\Fw(nT)-\Fs_1(nT)|.
\ee
This coincidence is  due  to the   fact  that the   interpolator  in \ref{eq:56} is   also
reconstructing the  noise realization $\Fw(t)$,  since it is  also a signal with bandwidth
$B$.
\begin{figure}
\begin{center}
\includegraphics{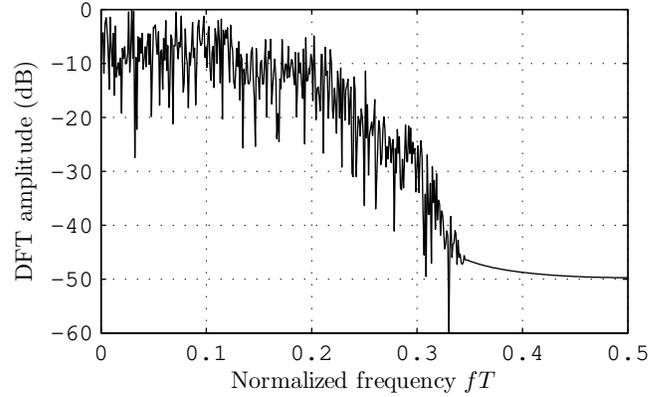}
\end{center}
\caption{\label{fig:8} Amplitude spectrum of sequence $\Fs(nT)+\Fw(nT)$ for $nT$ in
  $I$. The spectrum's maximum value has been normalized to one.}
\end{figure}
Fig. \ref{fig:8}
shows  the   amplitude spectrum  of   the sequence  $\Fs(nT)+\Fw(nT)$,  where  the maximum
has been  normalized to $0$ dB.   If this spectrum were  computed from the  sine
wave crossings, the it would differ from  that if Fig. \ref{fig:8}  by the amplitude given
in Fig. \ref{fig:9}.

\section{Conclusions}

A method has been presented that makes  it possible to  recover a band-limited signal from
its crossings with a sine wave. It allows one to design A/D  converters which only deliver
the timing  of the sine  wave crossings, so  allowing for a  smaller complexity  and power
consumption  in the converter.   The method is  based  on viewing  the  problem as one  of
interpolation from nonuniform  samples, to which a  recent efficient technique is applied.
This    technique  is based   on   the Lagrange  interpolator    and allows  for efficient
implementations based on current designs of VFD filters. As a spin-off, the method permits
one to perform spectral analysis from the sine  wave crossings with  the complexity of the
FFT.  The method has been validated in several numerical examples.

\appendices

\section{Weight function for the Lagrange interpolator}
\label{ap:wfl}

Following \cite{Selva09}, it is first  necessary to define  a band-limited window function
$\Fw(t)$ that approximately  selects a finite  time range.  This  function is  the inverse
Fourier transform of the Kaiser-Bessel window,
\be{eq:30}
\Fw(t)\equiv \frac{\sinc(B_w\sqrt{t^2-T_w^2})}{\sinc(j B_w T_w)},
\ee
where
\be{eq:31}
B_w\equiv 1/T-B \eqand T_w=PT.
\ee
Note that in (\ref{eq:30}) the argument of the sinc functions may be  pure
imaginary. In this case, the sinc can be  evaluated from the hyperbolic
sine since, for real $a$, it is 
\be{eq:32}
\sinc(ja)=\frac{\sin(j\pi a)}{j\pi a}=\frac{\Fe^{-\pi  a}-\Fe^{\pi  a}}{(2j)(j\pi a)}=
\frac{\sinh(\pi a)}{\pi a}.
\ee
The weight function $\Fgam(t)$ in Sec. \ref{sec:ps} is then given by
\be{eq:33}
\Fgam(t)\equiv\frac{(-1)^P}{(P!)^2}\frac{\Fw(t)\FL_o(t)}{\sin(\pi t/T)},
\ee
where $\FL_o(t)$ is the Lagrange kernel for the instants $pT$, $|p|\leq P$,
\be{eq:98}\nonumber
\FL_o(t)\equiv \prod_{p=-P}^P t-pT.
\ee

\bibliographystyle{IEEEbib}

\bibliography{../../../Utilities/LaTeX/Bibliography}

\end{document}